\documentclass[preprint,aps,superscriptaddress,showpacs,preprint,preprintnumbers,amsmath,amssymb,nofootinbib,tightenlines]{revtex4}

\usepackage{graphicx}
\usepackage{dcolumn}
\usepackage{bm}

\usepackage{color}

\def\mpi2{m_\pi^2}
\def\mK2{m_K^2}

\begin{document}
\bibliographystyle{apsrev}

\includegraphics[width=5cm]{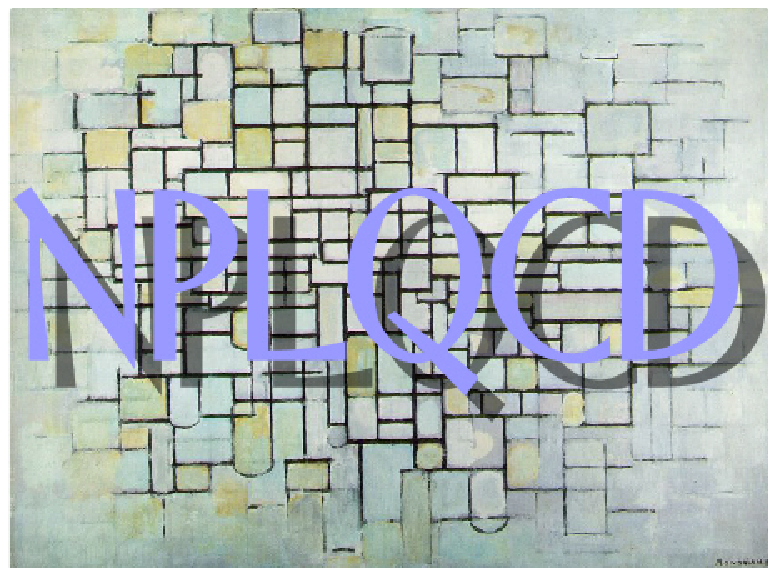}

\preprint{\vbox{
\hbox{NT@UW-06-15}
\hbox{JLAB-THY-06-503}
\hbox{UMD-40762-365}
\hbox{UNH-06-05}
}}

\vskip 0.9cm

\title{$f_K/f_\pi$ in Full QCD with Domain Wall Valence Quarks}

\author{S.R.~Beane}
\affiliation{Department of Physics, University of New Hampshire, Durham, NH 03824-3568}
\affiliation{Jefferson Laboratory, 12000 Jefferson Avenue, Newport News, VA 23606}

\author{P.F.~Bedaque}
\affiliation{Department of Physics, University of Maryland, College Park, MD 20742-4111}
\affiliation{Lawrence-Berkeley Laboratory, Berkeley, CA 94720.}

\author{K.~Orginos}
\affiliation{Department of Physics, College of William and Mary, Williamsburg, VA 23187-8795}
\affiliation{Jefferson Laboratory, 12000 Jefferson Avenue, Newport News, VA 23606}

\author{M.J.~Savage}
\affiliation{Department of Physics, University of Washington, Seattle, WA 98195-1560}
\collaboration{ NPLQCD Collaboration }

\date{\today}

\begin{abstract}
\noindent We compute the ratio of pseudoscalar decay constants
$f_K/f_\pi$ using domain-wall valence quarks and rooted improved
Kogut-Susskind sea quarks.  By employing continuum chiral perturbation
theory, we extract the Gasser-Leutwyler low-energy constant $L_5$, and
extrapolate $f_K/f_\pi$ to the physical point. We find: $f_K/f_\pi
= 1.218 \pm 0.002\ ^{+0.011}_{-0.024}$ where the first error is
statistical and the second error is an estimate of the systematic due to
chiral extrapolation and fitting procedures. This value
agrees within the uncertainties with the determination by the MILC
collaboration, calculated using Kogut-Susskind valence quarks,
indicating that systematic errors arising from the choice of lattice
valence quark are small.
\end{abstract}

\maketitle

\newpage


\section{Introduction}
\label{sec:intro}

\noindent 
Recently, lattice QCD calculations have been quite successful in
determining the hadronic matrix elements and low-energy constants
required for precisely extracting CKM matrix elements, such as $V_{bc}$
and $V_{us}$, from experimental
data~\cite{Aubin:2004fs,Bernard:2005ei,Marciano:2004uf,Okamoto:2005zg,Mackenzie:2005wu,El-Khadra:2001rv,Okamoto:2003ur}.
In particular, lattice determinations of the pseudoscalar decay
constants~\cite{Aubin:2004fs} $f_K$ and $f_\pi$, when combined with
the experimentally-measured branching fractions for
$K\rightarrow\mu\overline{\nu}_\mu (\gamma)$ and
$\pi\rightarrow\mu\overline{\nu}_\mu (\gamma)$, provide important
theoretical input into establishing the value of
$V_{us}$~\cite{Marciano:2004uf}, the charged-current matrix element
for $s\rightarrow u$ transitions.  Precise determinations of $V_{us}$
and $V_{ud}$, together with the fact that the square of $V_{ub}$ is negligibly
small, provide for a clean test of the unitarity of the CKM matrix,
and therefore facilitate a low-energy probe for physics beyond the
standard model with three generations of quarks.

Recent developments in improving the Kogut-Susskind (KS)
action~\cite{Lagae:1998he,Lagae:1998pe,Toussaint:1998sa,Orginos:1998ue,Lepage:1998vj,Orginos:1999cr}
have allowed for the computation of quantities in full QCD, with two
light and one strange dynamical quark flavors~\cite{Bernard:2001av},
near the physical point. Although such calculations currently
represent the most accurately calculated predictions of QCD, one
should keep in mind that there may be uncontrolled errors due to the
fact that KS fermions naturally appear with four copies (tastes).  In
order to use them in computations with one or two flavors one must
take fractional powers of the KS fermionic determinant, which may lead
to errors arising from non-localities.  While this problem  remains
under investigation, there exists significant evidence that, in
practice, this procedure is benign.  The low-energy effective field
theories describing quantities computed on the lattice with KS
fermions which are used to perform chiral and continuum
extrapolations, and also to determine finite-volume effects, are
complicated by the taste structure, which introduces new low-energy
constants~\cite{Aubin:2003mg,Aubin:2003uc} beyond those that appear in
the low-energy effective field theory of QCD.  Using the LHPC
mixed-action calculational scheme~\cite{Renner:2004ck,Edwards:2005kw},
one can alleviate the above-mentioned problems as flavor symmetry and
chiral symmetry (up to exponentially-small corrections) can be
preserved in the valence sector by the use of domain-wall
fermions~\cite{Kaplan:1992bt,Shamir:1993im,Shamir:1993zy,Shamir:1998ww,Furman:1995ky}.
Even in this scheme, the finite lattice spacing corrections due to the
sea of KS fermions are involved~\cite{Bar:2005tu,Chen:2005ab}, but
they are $O(g^2 b^2)$ (where $g$ is the QCD coupling constant and $b$
is the lattice spacing) and in some cases they may be negligible as
was observed in the case of I=2 $\pi\pi$
scattering~\cite{Beane:2005rj} and the more recent exploration of the
Gell-Mann-Okubo relation for octet baryons~\cite{Beane:2006fk}, and
calculation of the strong isospin breaking in the
nucleon~\cite{Beane:2006pt}.

In the calculation described here, we use the MILC rooted KS 2+1
dynamical fermion lattices~\cite{Orginos:1999cr,
Orginos:1998ue,Bernard:2001av,Bernard:2002pc} at a lattice spacing of
$b=0.125~{\rm fm}$ and domain-wall valence
quarks~\cite{Kaplan:1992bt,Shamir:1993im,Shamir:1993zy,Shamir:1998ww,Furman:1995ky}
to compute the pseudoscalar decay constants $f_K$ and $f_\pi$, and in
particular the ratio of the two.  As any deviation from unity in the
ratio of the decay constants results from the breaking of $SU(3)$
flavor symmetry, contributions from finite lattice spacing must be
accompanied by $SU(3)$ breaking quantities, and therefore are
suppressed beyond the naive $O(g^2 b^2)$.  It follows that it is
appropriate to employ continuum $SU(3)$ chiral perturbation theory to
extrapolate the lattice data to the physical values of the light-quark
masses, and to make a prediction for $f_K/f_\pi$.  This calculation
provides an important test of the systematics involved in the earlier
calculations of the same quantity by
MILC~\cite{Aubin:2004fs,Bernard:2005ei}.  Significant differences
between the two extrapolations would indicate an uncontrolled
systematic associated with the species of valence quarks employed in
the calculation.  In this paper we obtain a result that is consistent
with the MILC result, and consequently, we find no evidence of a
significant systematic error in the lattice calculation of $f_K/f_\pi$
due to finite lattice spacing effects.

\section{Details of the Lattice Calculation}
\label{sec:Latt}

\noindent 
Our computation uses the mixed-action lattice QCD scheme developed by
LHPC~\cite{Renner:2004ck,Edwards:2005kw} using domain-wall valence
quarks from a smeared-source on $N_f=2+1$
asqtad-improved~\cite{Orginos:1999cr,Orginos:1998ue} MILC
configurations generated with rooted~\footnote{For recent discussions
of the ``legality'' of the mixed-action and rooting procedures, see
Ref.~\cite{Durr:2004ta,Creutz:2006ys,Bernard:2006vv,Durr:2006ze,Hasenfratz:2006nw}.}
KS sea quarks~\cite{Bernard:2001av} that are hypercubic-smeared
(HYP-smeared)~\cite{Hasenfratz:2001hp,DeGrand:2002vu,DeGrand:2003in,Durr:2004as}.
In
the generation of the MILC configurations, the strange-quark mass was
fixed near its physical value, $b m_s = 0.050$, (where $b=0.125~{\rm
fm}$ is the lattice spacing~\footnote{The lattice spacing has been determined to be~\cite{Aubin:2004fs}
$b=0.1243\pm 0.0015~{\rm fm}$ using the Sommer scale-setting procedure, and~\cite{Beane:2005rj}
$b=0.1274\pm 0.0007\pm 0.0003~{\rm fm}$ using the pion decay constant. In this work quantities in physical units
were obtained using $b=0.125~{\rm fm}$.}) 
determined by the mass of hadrons containing strange quarks.  
The two light quarks in the configurations
are degenerate (isospin-symmetric).  
As was shown
by LHPC~\cite{Renner:2004ck,Edwards:2005kw}, HYP-smearing 
allows for a significant
reduction in the residual chiral symmetry breaking at a moderate
extent $L_s = 16$ of the extra dimension and domain-wall height
$M_5=1.7$.  
Using Dirichlet boundary conditions the
original time extent was reduced from  64 down to 32. This allowed us to recycle
propagators computed for the nucleon structure function calculations
performed by LHPC. For bare domain-wall fermion masses we used the
tuned values that match the KS Goldstone pion to
few-percent precision. For details of the matching
see Refs.~\cite{Renner:2004ck,Edwards:2005kw}.  The parameters used in the
propagator calculation are summarized in Table~\ref{tab:MILCcnfs}. All
propagator calculations were performed using the Chroma software
suite~\cite{Edwards:2004sx,McClendon:2001aa} on the high-performance
computing systems at the Jefferson Laboratory (JLab).
\begin{table}[tbp]
 \caption{The parameters of the MILC gauge configurations and
   domain-wall propagators used in this work. The subscript $l$
   denotes light quark (up and down), and  $s$ denotes the strange
   quark. The superscript $dwf$ denotes the bare quark mass for the
   domain wall fermion propagator calculation. The last column is the 
   number of configurations times the number of sources per
   configuration.}
\label{tab:MILCcnfs}
\begin{ruledtabular}
\begin{tabular}{ccccccc}
 Ensemble        
&  $b m_l$ &  $b m_s$ & $b m^{dwf}_l$ & $ b m^{dwf}_s $ & $10^3 \times b
m_{res}$~\protect\footnote{Computed by the LHP collaboration.} & \# of propagators  \\
\hline 
2064f21b676m007m050 &  0.007 & 0.050 & 0.0081 & 0.081  & $1.604\pm 0.038$ & 468$\times$3 \\
2064f21b676m010m050 &  0.010 & 0.050 & 0.0138 & 0.081  & $1.552\pm 0.027$ & 658$\times$4 \\
2064f21b679m020m050 &  0.020 & 0.050 & 0.0313 & 0.081  & $1.239\pm 0.028$ & 486$\times$3 \\
2064f21b681m030m050 &  0.030 & 0.050 & 0.0478 & 0.081  & $0.982\pm 0.030$ & 564$\times$3 \\
\end{tabular}
\end{ruledtabular}
\end{table}

In order to be able to extract the pseudoscalar decay constants from
the amplitude of the pseudoscalar correlators, $C_{P}(t)$, as was done
in~\cite{Blum:2000kn,Aoki:2002vt}, both the smeared-smeared and
smeared-point pseudoscalar correlation functions are computed.  If the
amplitudes of the pseudoscalar ground state are ${\cal A}_P^{ss}$ and
${\cal A}_P^{sp}$ for the smeared-smeared and smeared-point correlators,
respectively, the pseudoscalar decay constant is recovered from
\begin{equation}
  f_{P} = \frac{{\cal A}_P^{sp}}{\sqrt{{\cal A}_P^{ss}}} 
\frac{\sqrt{2}(m^{dwf}_1 + m^{dwf}_2  + 2 m_{res})}{m_{P}^{3/2}}
\label{eq:Fps}
\end{equation}
where $m^{dwf}_1$ and $m^{dwf}_2$ are the domain-wall fermion masses
used in constructing the pseudoscalar meson and $m_{res}$ is the
residual chiral symmetry breaking parameter computed from the chiral
Ward-Takahashi identity as in~\cite{Blum:2000kn,Aoki:2002vt} and shown
in Table~\ref{tab:MILCcnfs}. The dependence of $m_{res}$ on the
valence mass is negligible compared to the statistical errors of the
calculation. 
It is useful to  construct an ``effective'' decay constant
directly from the lattice data at each time slice. Hence we form
\begin{equation}
  f^{EFF}_{P} \ =\  
\frac{C^{SP}_{P}(t)^{t+1}\, C^{SS}_{P}(t+1)^{t/2}}{C^{SS}_{P}(t)^{(t+1)/2}\,  C^{SP}_{P}(t+1)^{t}}
\ \frac{\sqrt{2}(m^{dwf}_1 + m^{dwf}_2  + 2 m_{res})}{\lbrack\, 
\log\left(C^{SP}_{P}(t) C^{SP}_{P}(t+1)^{-1}\right) \, \rbrack^{3/2}} \ ,
\label{eq:FpsEFF}
\end{equation}
which is independent of $t$ at large times where the correlation functions behave as
\begin{eqnarray}
C^{SP}_{P}(t) & & \rightarrow \ {\cal A}_P^{sp}\ e^{-m_P t}
\ \ \ \ , \ \ \ \ 
C^{SS}_{P}(t) \ \rightarrow \ {\cal A}_P^{ss}\ e^{-m_P t}
\ \ \ .
\end{eqnarray}

\section{Analysis and Chiral Extrapolation}

\noindent To determine the pseudoscalar decay constants, the correlation functions for
the $K^+$ and $\pi^+$ were computed with both smeared and point sinks on each
ensemble. 
%
\begin{figure}[!ht]
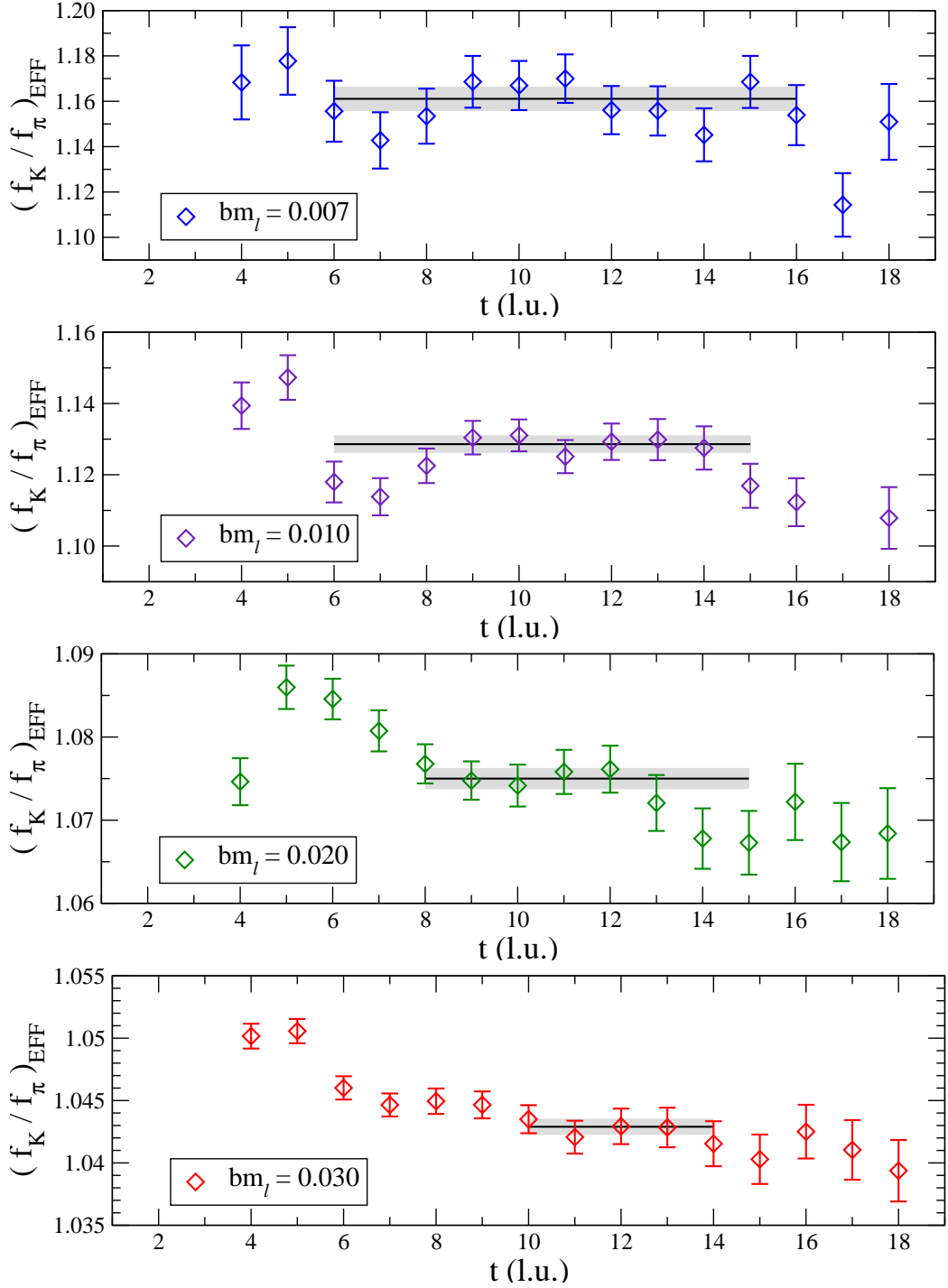

\centering
\includegraphics*[width=0.85\textwidth,viewport=2 5 700 240,clip]{fkfpi_EFF007.eps}
\hfill
\includegraphics*[width=0.85\textwidth,viewport=2 5 700 240,clip]{fkfpi_EFF010.eps}
\hfill
\includegraphics*[width=0.85\textwidth,viewport=2 5 700 240,clip]{fkfpi_EFF020.eps}
\hfill
\includegraphics*[width=0.85\textwidth,viewport=2 5 700 240,clip]{fkfpi_EFF030.eps}
\caption{\it ``Effective'' $f_K/f_\pi$ determined 
from the smeared-smeared and smeared-point correlation functions
with eq.~(\protect\ref{eq:FpsEFF}).
The solid
black lines and shaded regions are the fits (with $1\sigma$ errors) 
tabulated in Table~\protect\ref{tab:LatResults}.}
\label{fig:SSSPcorplots}
\end{figure}
In order to extract the amplitudes for the smeared-smeared and smeared-point
correlation functions a single exponential with a common mass was fit 
by $\chi^2$-minimization to each
data set, i.e. a three parameter fit was performed with variables
${\cal A}_P^{ss}$, ${\cal A}_P^{sp}$ and $m_\pi$ (or $m_K$).
The central value and uncertainty of each parameter was determined by the
jackknife procedure over the ensemble of configurations.
The decay constant was extracted by jackknifing over the appropriate combination
of quantities, as given in eq.~(\ref{eq:Fps}). In fig.~\ref{fig:SSSPcorplots} 
we present the lattice data using ``effective'' $f_K/f_\pi$ plots according to 
eq.~(\protect\ref{eq:FpsEFF}), along with the fits.
\begin{table}[!ht]
 \caption{Calculated quantities with fitting ranges in brackets. All errors are computed using
   the jackknife procedure. 
Values for ${\cal F}$ are given
without and with (in squiggly brackets) the $log^2$ contribution.}
\label{tab:LatResults}
\begin{ruledtabular}
\begin{tabular}{cccccc}
 Ensemble        
&  $m_\pi$(GeV) & $m_{\pi}/f_{\pi}$ & $m_K/f_\pi$ &  $f_K/f_{\pi}$ & ${\cal F}\times 10^3$\\
\hline 
m007 &0.2931(15)& 1.978(19) [5-16]  & 3.937(28) [6-16] & 1.1610(54) [6-16]&  5.67(6) \{5.31(5)\}      \\
m010 & 0.3546(9)& 2.337(11) [5-16] & 3.958(16) [6-15]& 1.1286(23) [6-15] &  5.62(3) \{5.13(2)\}     \\
m020 & 0.4934(12)& 3.059(12) [7-16]  & 3.988(15) [7-15] & 1.0751(13) [8-15]  &  5.68(3) \{4.87(2)\}     \\
m030 & 0.5918(10)& 3.484(10) [5-15]  & 4.004(12) [7-14] & 1.04279(69) [10-14] & 5.73(2) \{4.69(2)\}    
\end{tabular}
\end{ruledtabular}
\end{table}
The results of the lattice calculation of the 
decay constants and meson masses
are tabulated in Table~\ref{tab:LatResults}.

\subsection{Chiral Extrapolation at Next-to-Leading Order}
\label{sec:ChiPT}

\noindent 
In $SU(3)$ chiral perturbation theory ($\chi$PT) Gasser and
Leutwyler~\cite{Gasser:1984gg,Gasser:1983yg,Gasser:1983ky} showed
that the ratio of the kaon to pion decay constants is given, at 
next-to-leading order (NLO) in the chiral expansion, by
\begin{equation}
\frac{f_K}{f_\pi} = 1 + \frac{5}{4}  l_\pi(\mu) -  \frac{1}{2}  l_K(\mu) -   \frac{3}{4}  l_\eta(\mu) 
+ \frac{8}{f^2} \left(m^2_K - m^2_\pi\right) L_5(\mu)
\label{eq:GLfkofpi}
\end{equation}
where $f$ is the pseudoscalar decay constant in the chiral limit,
$m_K$ is the kaon mass, $m_\pi$ is the pion mass, and
\begin{equation}
l_i(\mu) \equiv \frac{1}{16 \pi^2 } \frac{m_i^2}{f^2}\log\left( \frac{m_i^2}{\mu^2}\right),
\end{equation}
with the index $i$ running over the pseudoscalar states ($\pi$,$K$ and
$\eta$). 
$L_5(\mu)$ is a Gasser-Leutwyler low-energy constant
evaluated at the $\chi$PT renormalization scale $\mu$, whose scale dependence
exactly compensates the scale dependence of the logarithmic contributions.

In our lattice calculation we have not computed the mass of the $\eta$
meson since it involves disconnected diagrams that require significant 
computer time to evaluate.  Hence we replace $m_\eta$ with
its value obtained from the Gell-Mann-Okubo mass-relation among octet mesons,
\begin{equation}
m_\eta^2 = \frac{4}{3}m_K^2 -  \frac{1}{3}m_\pi^2
\ \ ,
\end{equation}
which is valid to the order of $\chi$PT to which we are working. In addition,
we choose to work with $\mu=f_\pi^{\it phy}$, the value of the pion decay constant at
the physical point.  
To recover the value of the counterterm $L_5(\mu)$ at some other
renormalization scale, one can use the evolution~\cite{Gasser:1984gg,Gasser:1983yg,Gasser:1983ky}
\begin{equation}
L_5(f_\pi^{\it phy})  = L_5(\mu) - \frac{3}{8}  \frac{1}{16 \pi^2 } 
\log\left( \frac{f_\pi^{\it phy}}{\mu}\right) \ .
\end{equation}
Finally, we replace the ratios $(m_i/f_\pi^{\it phy})^2$ by the
lattice-computed value $(m_i/f_\pi)^2$, which is again consistent to
the order of $\chi$PT to which we are working. 
Hence, the final NLO expression to which we fit the  lattice data is
\begin{eqnarray}
  \frac{f_K}{f_\pi} = 1 
    &+& \frac{5}{4}\frac{1}{16 \pi^2 }\frac{m_\pi^2}{f_\pi^2}\log\left( \frac{m_\pi^2}{f_\pi^2}\right)    
    -  \frac{1}{2}\frac{1}{16 \pi^2 }\frac{m_K^2}{f_\pi^2}\log\left( \frac{m_K^2}{f_\pi^2}\right) \nonumber\\
    &-&  \frac{1}{16 \pi^2 }\left( \frac{m_K^2}{f_\pi^2} - \frac{1}{4} \frac{m_\pi^2}{f_\pi^2} \right)
 \log\left( \frac{4}{3} \frac{m_K^2}{f_\pi^2} - \frac{1}{3} \frac{m_\pi^2}{f_\pi^2} \right)
    + 8 \left( \frac{m_K^2}{f_\pi^2}   - \frac{m_\pi^2}{f_\pi^2}   \right) L_5(f_\pi^{\it phy}) \ .
    \label{eq:Latfkofpi}
 \end{eqnarray}
Note that the only parameter to be determined by fitting at
NLO is $L_5$.  It is also worth noting that the above expression has the
expected behavior that at the $SU(3)$ symmetric point the ratio
of decay constants is unity. 

For reasons that will become clear below, it is useful
to ``linearize'' the fitting procedure by isolating the analytic terms with
coefficients that are to be fit to the lattice data.
We define the function
\begin{equation}
{\cal F} \equiv \left(\frac{f_K}{f_\pi} -  1 - \chi_{logs}\right)
\frac{1}{8 \ y} 
\ ,
\label{eq:Fdef}
\end{equation}
where, at NLO,
\begin{eqnarray}
  \chi_{logs}\ =\  \chi_{logs}^{(NLO)}\left(\frac{m_\pi}{f_\pi}, \frac{m_K}{f_\pi}\right) =  
    & & \frac{5}{4}\frac{1}{16 \pi^2 }\frac{m_\pi^2}{f_\pi^2}\log\left( \frac{m_\pi^2}{f_\pi^2}\right) 
    -  \frac{1}{2}\frac{1}{16 \pi^2 }\frac{m_K^2}{f_\pi^2}\log\left( \frac{m_K^2}{f_\pi^2}\right) \nonumber\\
    & & -  \frac{1}{16 \pi^2 }\left( \frac{m_K^2}{f_\pi^2} - \frac{1}{4} \frac{m_\pi^2}{f_\pi^2} \right)
 \log\left( \frac{4}{3} \frac{m_K^2}{f_\pi^2} - \frac{1}{3}
   \frac{m_\pi^2}{f_\pi^2} \right) 
\ \ \ ,
\label{eq:chidefs}
\end{eqnarray}
and the quantity $y$ is 
\begin{equation}
  y = \frac{m^2_K}{f_\pi^2} - \frac{m^2_\pi}{f_\pi^2} 
\ .
\end{equation}
Therefore, at NLO in the chiral expansion, the quantity 
${\cal F}$ should be the same on each of the ensembles, and equal to the
counterterm 
$L_5(f_\pi^{\it phy})$,
\begin{eqnarray}
{\cal F} & = & L_5(f_\pi^{\it phy})
\ \ \ .
\label{eq:FNLO}
\end{eqnarray}
\begin{figure}
\begin{center}
\includegraphics[width=0.8\textwidth]{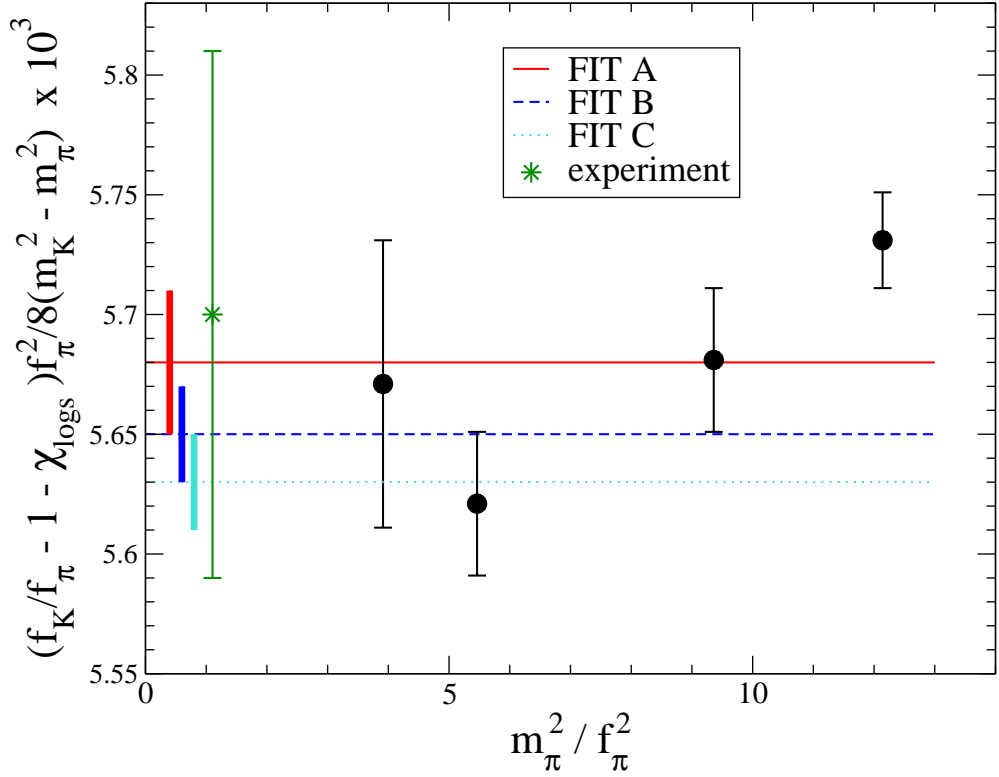}
\end{center}
\caption{
${\cal F}$ vs. $m^2_{\pi}/f^2_{\pi}$ at NLO, along with the three
different fits, A, B and C.
The solid bars near the y-axis denote the value of 
$L_5$ and its uncertainty from the three fits.
The point denoted by the star corresponds to the experimental value.
}
\label{fig:LatFitxNLO}
\end{figure}
The calculated values of ${\cal F}$, along with their uncertainties
determined by jackknifing over the configurations, are shown in
Table~\ref{tab:LatResults}, and in fig.~\ref{fig:LatFitxNLO} we have
plotted ${\cal F}$ versus $m_\pi^2/f_\pi^2$.  A
$\chi^2$-minimization is performed to extract the one parameter
$L_5(f_\pi^{\it phy})$ from the data.  It is clear that the data is
not that well fit by a constant, due to the presence of higher-order
terms in the chiral expansion, and so to explore the dependence on
these higher order terms we have sequentially ``pruned'' the data by
removing the highest mass point ($b m_l=030$), and then the two
highest mass points ($b m_l=030$, $020$) and determined
$L_5(f_\pi^{\it phy})$~\footnote{ Pruning the data provides an
assessment of the importance of higher-order terms in the chiral
expansion, while fitting only the leading chiral contributions.  There
are a number of ways to approach this issue.  For instance, an
alternate approach would be to add a systematic error to each data
point that grows with the pion mass in a manner consistent with $\chi$PT.
We find that this provides an extrapolated value of $f_K/f_\pi$ and
$L_5$ consistent with pruning the data, as expected.  }.  The results
of these fits are shown in fig.~\ref{fig:LatFitxNLO}, and presented in
Table~\ref{tab:FitResultsNLO}.
\begin{table}[!ht]
 \caption{Results from chiral extrapolation at one-loop order in $\chi$PT. 
Explanation of the various fits is in the text.}
\label{tab:FitResultsNLO}
\begin{ruledtabular}
\begin{tabular}{cccc}
FIT &  $L_5\times 10^3$ &   $f_K/f_{\pi}$ (extrapolated) & $\chi^2$/dof \\
\hline 
A & $5.68(3)$  & $1.221(3)$ &  $3.5$  \\
B & $5.65(2)$  & $1.218(2)$ &  $1.4$  \\
C & $5.63(2)$  & $1.215(2)$ &  $0.7$  \\
\end{tabular}
\end{ruledtabular}
\end{table}
With the value of $L_5$, we use eq.~\ref{eq:Latfkofpi} to
evaluate the ratio of the decay constants at the physical point using
the physical values for the pseudoscalar masses and the pion decay
constant~\cite{Eidelman:2004wy},
\begin{eqnarray}
  f_{\pi^+} &=& 130.7~{\rm MeV} \ ,\  m_\pi \ = \ 137.3~{\rm MeV} \ ,\  m_K   \ =\  495.7~{\rm MeV} ,
\end{eqnarray}
where the masses are the isospin-averaged values.  We use the
Gell-Mann--Okubo mass relation to determine the $\eta$-mass that appears
in the chiral contributions.

It is important to keep in mind that this determination of $L_5$  is only
perturbatively close to the actual value of $L_5$ which is defined in the
chiral limit.  In the current extraction, the strange quark mass is held fixed near the
physical value, while the light quark masses are somewhat lighter.

\subsection{Incomplete Chiral Extrapolations at Next-to-Next-to-Leading Order}

\noindent While the full two-loop expressions for ${f_K}/{f_\pi}$
exist in both QCD~\cite{Amoros:1999dp} and partially-quenched
QCD~\cite{Bijnens:2006jv}, these expressions contain many fit
parameters, and therefore fruitful use of these results must await
lattice data with better statistics and at a larger variety of quark
masses. In order to estimate systematic errors, we perform fits with
parts of the next-to-next-to-Leading-Order (NNLO)
expression~\cite{Bijnens:1998yu}.  We focus on just two of the
structures that enter at NNLO, analytic terms and a double logarithm
with fixed coefficient.

\subsubsection{Partial $N^2LO$ : Analytic Terms Only}

\begin{figure}
\vskip.4in
\begin{center}
\includegraphics[width=0.8\textwidth]{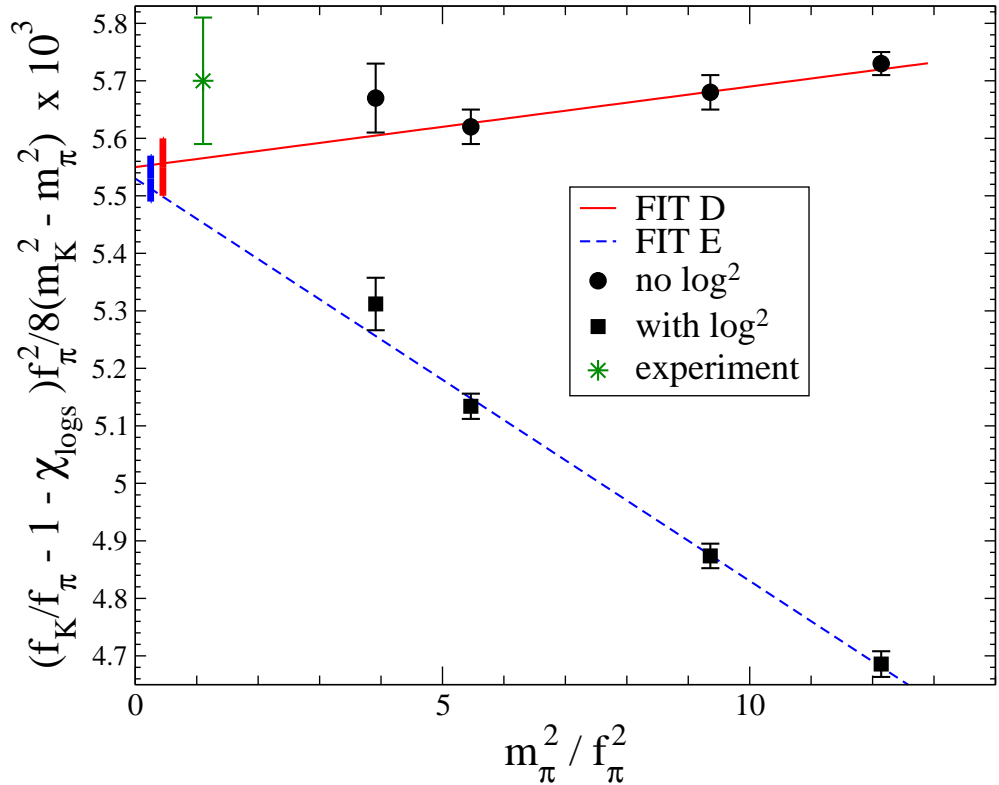}
\end{center}
\caption{${\cal F}$ vs. $m^2_{\pi}/f^2_{\pi}$ at NNLO.
The solid bars near the y-axis denote the value of 
$\tilde L_5={\cal F}(m_\pi =0)$ and its uncertainty from the fits.
The point denoted by the star corresponds to the experimental value.
The circles denote the lattice data with only the NLO chiral logs subtracted,
while the squares are the lattice data with the NLO chiral logs and the
NNLO $\log^2$ term subtracted.
}
\label{fig:LatFitxNNLO}
\end{figure}

\begin{figure}
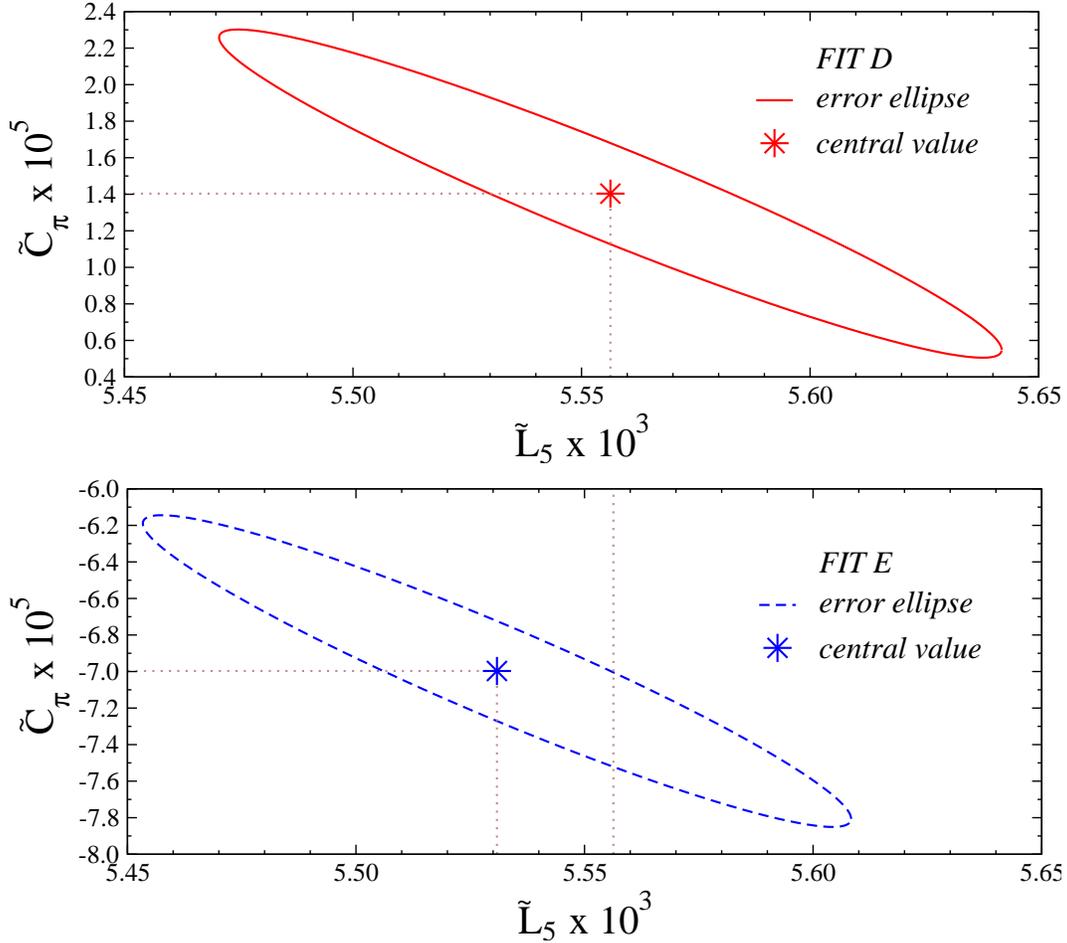

\begin{center}
\includegraphics*[width=0.85\textwidth,viewport=2 2 706 320,clip]{ellipse_1sig_D.eps}
\hfill
\includegraphics*[width=0.85\textwidth,viewport=2 2 706 320,clip]{ellipse_1sig_E.eps}
\end{center}
\caption{68\% confidence-level error ellipses for fits D and E described in the text.}
\label{fig:errorellipse}
\end{figure}

\noindent Including only the analytic terms that enter at NNLO, eq.~(\ref{eq:FNLO}) becomes
\begin{eqnarray}
{\cal F} & = & L_5\ +\ C_s \ m_s\ +\ C_l \ m_l
\nonumber\\
& = & \tilde L_5\ +\ \tilde C_\pi\ \frac{m_\pi^2}{f_\pi^2}
\ \ \ ,
\label{eq:FNNLOana}
\end{eqnarray}
where terms higher order in the chiral expansion are not shown.  As
the strange quark mass is the same over all ensembles, we simply
absorb it into the definition of $L_5$, making explicit the quark mass
dependence discussed previously.  Therefore fitting at NNLO holding
the strange quark mass fixed introduces one additional fit parameter,
$\tilde C_\pi$.  It is clear that the values of $\tilde L_5$ and
$\tilde C_\pi$ extracted from the data are correlated, and in
determining the extrapolated value of $f_K/f_\pi$ we explore the
entire 68\% confidence-level error ellipse in the $\tilde L_5-\tilde C_\pi$ 
plane~\footnote{This results in an error that is consistent with 
textbook propogation of the errors in $\tilde L_5$ and $\tilde C_\pi$.}
(shown in fig.~\ref{fig:errorellipse}). We label this fit D, and 
the results are shown in Table~\ref{tab:FitResultsNNLO}.
\begin{table}[!ht]
 \caption{Results from fitting the partial NNLO chiral contributions. 
Explanation of fits D,E,F and G are in the text.}
\label{tab:FitResultsNNLO}
\begin{ruledtabular}
\begin{tabular}{ccccc}
FIT &  ${\tilde L}_5\times 10^3$ & $\tilde C_\pi\times 10^5$ &  $f_K/f_{\pi}$
(extrapolated) &
$\chi^2$/dof \\
\hline
D & $5.55(5)$  & $\;\;1.40(49)$  & $1.209(8)$ &  $0.8$  \\
E & $5.53(4)$  & $-7.00(47)$  & $1.209(7)$ &  $1.0$  \\ 
\hline
F & $5.80(3)$  & $-8.28(35)$  & $1.224(5)$ &  $0.5$  \\ 
G & $5.16(5)$  & $-4.93(56)$  & $1.205(9)$ &  $1.5$  \\ 
\end{tabular}
\end{ruledtabular}
\end{table}
The data minus the NLO chiral logs, and the fit are shown in fig.~\ref{fig:LatFitxNNLO}.
Note that the errors quoted in Table~\ref{tab:FitResultsNNLO}
and displayed in fig.~\ref{fig:LatFitxNNLO} are $1\sigma$ 
errors.
%
\subsubsection{Partial $N^2LO$ : Analytic Terms and Double Chiral Logs}

\noindent The full two-loop expression that contributes to $f_K/f_\pi$
is quite complicated.  An approximation to the $\log^2$ piece at
two-loop order can be evaluated using renormalization-group techniques
and is given by~\cite{Bijnens:1998yu}
\begin{eqnarray}
\chi_{\log^2}^{(NNLO)}& =&
\frac{1}{6144 \pi^4} \left(\frac{m^2_K}{f_\pi^2} - 
\frac{m^2_\pi}{f_\pi^2} \right) \left(17 \frac{m^2_K}{f_\pi^2} + 37 \frac{m^2_\pi}{f_\pi^2} \right)
\log\left( \frac{M^2}{\mu^2}\right)^2 
\ \ , 
\label{eq:doublelog} 
\end{eqnarray}
where $M$ is a mass scale related to the Goldstone boson masses. It would
seem reasonable to choose the intermediate mass scale $M=\sqrt{m_\pi m_K}$ and
$\mu=f_\pi$.  Of course, the two-loop contributions vanish at the
flavor $SU(3)$ symmetric point.  Again, to isolate the fitting function,
we subtract $\chi_{\log^2}^{(NNLO)}$ from the lattice data, giving a
fit function of the form
\begin{equation}
{\cal F} \equiv \left(\frac{f_K}{f_\pi} -  1 - \chi_{logs}^{(NLO)} -  \chi_{\log^2}^{(NNLO)}\right)
\frac{1}{8 \ y} 
\ =\ 
\tilde L_5(f_\pi^{\it phy})\ +\ \tilde C_\pi\ \frac{m_\pi^2}{f_\pi^2}
\ .
\label{eq:FdefNNLO}
\end{equation}
The scale dependence of the $\log^2$ contribution requires that the
coefficients $C_s$ and $C_u$ in eq.~(\ref{eq:FNNLOana}) be
scale-dependent, and thus $\tilde C_\pi$ becomes scale-dependent and
the scale-dependence of $\tilde L_5$ is modified (a higher order
effect).  The calculated values of ${\cal F}$, along with their
uncertainties determined by jackknifing over the configurations, are
shown in Table~\ref{tab:LatResults}, and plotted in
fig.~\ref{fig:LatFitxNNLO}. We anticipate that the fit value of
$\tilde L_5$ should change only a small amount from its value obtained
in the NLO fits and in fit D, if the chiral expansion is
convergent. However, we expect that the coefficient $\tilde C_\pi$
could change by an amount of order one.  The results of fitting this functional
form to the lattice data, which we denote by fit E, are presented in
Table~\ref{tab:FitResultsNNLO}, and shown in
fig.~\ref{fig:LatFitxNNLO} (error ellipse is shown in fig.~\ref{fig:errorellipse}).
Indeed, $\tilde L_5$ is changed very
little, while $\tilde C_\pi$ changes by an amount of order one.
We also give results for fits F and G which are the same as fit E
except the argument of the $log^2$ contribution is chosen to be
$M=m_\pi$ and $M=m_K$, respectively. These choices lead to a larger
variation in $\tilde L_5$ and consequently in $f_K/ f_\pi$.

\subsection{Discussion}
\label{sec:discuss}

\noindent To determine $f_K/ f_\pi$ at the physical point and its
associated uncertainty we synthesize the results of the NLO and NNLO
fits.  Fitting the lowest two mass points at NLO gives $f_K/ f_\pi =
1.215\pm 0.002$, while fitting the three data points with pion masses
below $m_\pi\sim 500~{\rm MeV}$ gives $f_K/ f_\pi = 1.218\pm 0.002$.
The difference between them is within statistical errors ($1.5
\sigma$) but there appears to be a systematic trend in the data which
can be attributed to higher orders in the chiral expansion. As we are
unable to fit the full NNLO expression to our small data set we can
estimate the systematic uncertainty in this calculation by looking at
the range of values of $f_K/ f_\pi$ that result from the two types of
NNLO extrapolation, both with and without the $\log^2$ contribution,
including variation in the argument of the NNLO logarithm and
including statistical errors.  The range of variation in the NNLO
estimate is an order of magnitude larger than the statistical error
found at NLO.  We take this NNLO uncertainty, $\Delta(f_K/ f_\pi)=^{+0.011}_{-0.022}$, to
be an estimate of the systematic error in our calculation due to the
truncation of the chiral expansion. We also assign a systematic error
due to fitting procedures, obtained by varying the fitting ranges
displayed in fig.~\ref{fig:SSSPcorplots}, which gives
$\Delta(f_K/ f_\pi)=^{+0.000}_{-0.010}$. Therefore, our final number is:
\begin{equation}
  \frac{f_K}{f_\pi} = 1.218 \pm 0.002\  ^{+0.011}_{-0.024}
\ ,
\end{equation}
where the first error is statistical and the second error is systematic,
with the extrapolation error and fitting error added in quadrature.
The error in this lattice QCD determination of ${f_K}/{f_\pi}$ is clearly 
dominated by the systematics.

Using a similar procedure, we arrive at a value for $L_5$:
\begin{eqnarray}
L_5(f_\pi^{\it phy}) &=& 5.65 \pm 0.02 \ ^{+0.18}_{-0.54} \ \times \ 10^{-3}
\ \ ,
\end{eqnarray}
where the first error is statistical and the second is an estimate of
the systematic error due to omitted higher orders in the chiral expansion. 
This then scales to give
$L_5(m_\eta^{\it phy}) = 2.22  \pm 0.02 \ ^{+0.18}_{-0.54}\,\times \, 10^{-3}$
at the $\eta$-mass 
and 
$L_5(m_\rho^{\it phy}) = 1.42   \pm 0.02 \ ^{+0.18}_{-0.54}\, \times \, 10^{-3}$ at the
$\rho$-mass.
As stated previously, this is an effective $L_5$ as it includes the 
higher order strange quark contribution.

The results for $f_K/f_\pi$ have an additional systematic error due to
the non-zero lattice spacing which we expect to be $O( (m_s-m_u) b^2
)$. In principle one can reduce this error by fitting to the
appropriate $\chi$PT formulas that include the $O(g^2 b^2)$ effects
due to flavor-symmetry breaking in the sea-quark
sector~\cite{Bar:2005tu}. However, our data fit well to the continuum
$\chi$PT formulas and hence we do not expect that use of the extended
$\chi$PT formulas of Ref.~\cite{Bar:2005tu} would significantly
improve our results at this stage. Our final result is consistent with
the MILC number~\cite{Aubin:2004fs}
\begin{equation}
  \left.\frac{f_K}{f_\pi}\right|_{\rm MILC} = 1.210 \pm 0.004 \pm 0.013 \ ,
\end{equation}
where the first error is statistical and the second is the total
systematic error estimated by MILC. Since our valence quarks are
domain-wall fermions, in contrast with the KS quarks used
by MILC, the discretization errors should be different.  Hence,
the agreement of our results to the few-percent level is further
confirmation that these systematic errors are small~\footnote{A more
recent MILC calculation~\cite{Bernard:2005ei} quotes a value of
$1.198 \pm 0.003\ ^{+0.016}_{-0.005}$. This calculation made use of finer lattices as
well as a second run with a lighter strange-quark mass at $b=0.125~{\rm
fm}$.}.

It is also interesting to note that our result is in agreement with
the experimental number,
\begin{equation}
  \left.\frac{f_K}{f_\pi}\right|_{\rm EXP} = 1.223 \pm 0.012
\ \ ,
\end{equation}
but our calculation has a  somewhat larger systematic error due to uncertainty in
the chiral extrapolation.

It is possible to further improve the precision of our calculation by
increasing the statistics of the lighter pion masses, including one
more point at even lighter pion mass, and by better utilizing the power of partial-quenching;
i.e. by computing with different valence quark masses, away from the
tuned point.
We hope that with these improvements in place we will be able to 
improve upon the MILC result.

\section{Conclusions}
\label{sec:Conc}

\noindent 
Existing high-precision calculations of basic standard model
quantities involve staggered valence quarks on staggered sea quarks
with their associated systematic errors. 
Clearly, it is  important to employ a variety of fermion discretizations in order 
to understand and reduce one of  the inherent systematic errors in 
lattice QCD calculations.
We have computed $f_K/f_\pi$
with domain-wall valence quarks on MILC lattices and find results
consistent with an earlier calculation by MILC using KS
valence quarks. It is gratifying to find that using different fermions in the
valence sector leads to a consistent precision determination of
$f_K/f_\pi$ in accord with basic effective field theory expectations
about the scaling of discretization errors.

\begin{acknowledgments}
\noindent This work was performed under the auspices of SciDAC.  We
thank R.~Edwards for help with the QDP++/Chroma programming
environment~\cite{Edwards:2004sx} with which the calculations
discussed here were performed. We are also indebted to the MILC and
the LHP collaborations for use of some of their configurations and
propagators, respectively.  This work was supported in part by the
U.S.~Dept.~of Energy under Grants No.~DE-FG03-97ER4014 (MJS),
No.~DF-FC02-94ER40818 (KO), No.~ER-40762-365  (PFB), the National
Science Foundation under grant No.~PHY-0400231 (SRB) and by DOE
through contract DE-AC05-84ER40150, under which the Southeastern
Universities Research Association (SURA) operates the Thomas Jefferson
National Accelerator Facility (KO,SRB).

\end{acknowledgments}




\bibliography{fkfpi_v1}

\end{document}